\documentclass[english]{revtex4}
\usepackage[T1]{fontenc}
\usepackage{amsmath}
\usepackage{amssymb}

\makeatletter
\@ifundefined{textcolor}{}
{%
 \definecolor{BLACK}{gray}{0}
 \definecolor{WHITE}{gray}{1}
 \definecolor{RED}{rgb}{1,0,0}
 \definecolor{GREEN}{rgb}{0,1,0}
 \definecolor{BLUE}{rgb}{0,0,1}
 \definecolor{CYAN}{cmyk}{1,0,0,0}
 \definecolor{MAGENTA}{cmyk}{0,1,0,0}
 \definecolor{YELLOW}{cmyk}{0,0,1,0}
}

\makeatother

\usepackage{babel}
\begin{document}
\title{MOND from Second-Order Moment Modified Acceleration and Quantum Equivalence
Principle}
\author{M.J.Luo}
\address{Department of Physics, Jiangsu University, Zhenjiang 212013, People's
Republic of China}
\email{mjluo@ujs.edu.cn}

\begin{abstract}
This paper proposes a novel non-inertial quantum effect wherein particle
spectra show second-order moment extra Gaussian broadening due to
local short-time (non-uniform) acceleration, as well as in a deSitter
spacetime background. Although the effect is too small to be detected,
it provides a mechanism for the cosmological constant to enter the
local kinematics of particles in the form of acceleration. The acceleration
composition relation of a proper motion acceleration and the cosmological
constant playing the role of a background acceleration, which is required
in the Modified Newtonian Dynamics (MOND). The origin of acceleration
discrepancies lies in the importance of the intrinsic second moment
quantum fluctuations in the deSitter background, so that the mean
value of derivative (modified effective acceleration) does not equal
to the derivative of mean value (the first moment acceleration given
by the unmodified Newtonian gravity). Such an interpretation of MOND
as a second moment effect necessitates a quantum equivalence principle
as its physical foundation, that is, extending the classical equivalence
at the level of mean values (first-order moments) to the quantum equivalence
at the level of second moment quantum fluctuations. The effective
distance quadratic form, effective curvature and effective acceleration,
etc., modified by the universal second moments all behave as if they
were real geometrical or physical quantities. This effect also offers
a unified framework for understanding the accelerated expansion of
the universe and the anomalies in galactic rotation curves or radial
acceleration.
\end{abstract}
\maketitle
\tableofcontents{}

\section{Introduction}

The theory of MOdified Newtonian Dynamics (MOND) \citep{1983A,Milgrom:2014usa,Milgrom:2016scu}
has achieved considerable success \citep{Famaey:2011kh,Milgrom:2014uwk,Milgrom:2019cle,Famaey:2025rma}
in replacing dark matter and explaining the anomalies in galactic
rotation curves and radial acceleration \citep{McGaugh:2016leg},
while the dark matter theory is seen less natural in achieving the
similar success. Nevertheless, MOND theory also faces numerous challenges
of its own, with the most fundamental one being the lack of a relativistic
covariant theoretical foundation. This manifests in several ways.
Firstly, the interpolation function used in MOND to describe the transition
from Newtonian dynamics to the deep-MOND regime (where there is a
significant deviation from Newtonian gravity) cannot be derived from
first principles. Consequently, there exist various forms of this
function that can roughly fit observational data in a phenomenological
manner. A fundamental theory underlying MOND must first be able to
derive this interpolation function.

Secondly, the interpolation function in MOND relies on the ratio between
the system's acceleration and a critical acceleration constant. Since
acceleration is a coordinate-dependent concept, MOND is not a covariant
theory because such a theory is tied to the choice of coordinate system.
In other words, this theory fails to address the question of which
frame the acceleration is being referred to. In attempts like Bimetric
MOND \citep{Milgrom:2009gv,Milgrom:2013jma,Milgrom:2022nyl}, Milgrom
has sought to construct a concept analogous to a covariant acceleration
vector by taking the difference in connections between two distinct
spacetime geometries---that is, treating this acceleration as the
acceleration of the geometry of the real spacetime relative to an
auxiliary geometry. However, these attempts have incurred certain
costs to achieve the specific objectives.

Thirdly, the behavior of MOND can be interpreted in two seemingly
equivalent ways. One approach attributes the interpolation function
to modifications of the gravitational theory, interpreting it as alterations
to gravitational potential, the Newton-Poisson action, or the action
of general relativity (Modified Gravity). Most well-developed MOND
theories proposed to date fall into this category, such as Bekenstein-Milgrom
MOND \citep{Bekenstein:1984tv}, quasi-linear MOND \citep{Milgrom:2009ee},
relativistic MOND \citep{Bekenstein:2004ne,Skordis:2009bf,Zlosnik:2006sb},
and Bimetric MOND \citep{Milgrom:2009gv,Milgrom:2013jma,Milgrom:2022nyl},
among others. The alternative approach keeps gravity unchanged but
attributes the interpolation function to modifications of Newton's
second law or the kinematics of free test particles (Modified Inertial)
\citep{Milgrom:1992hr,Milgrom:2011kx,Milgrom:2014uwk,Milgrom:2022ifm,Milgrom:2023pmv,Namouni:2015pga,2017,Costa:2019pbz}.
The behavior required by MOND imposes strong constraints on the mathematical
form of metric-based gravitational theories. As we know, covariance
also imposes strong constraints on the possible forms of gravity,
to the extent that reconciling both sets of constraints naturally
can be challenging. For instance, generally covariant theories like
$f(R)$ are unable to produce MOND-like behavior \citep{Milgrom:2019rtd}
(except for some radically modified forms, such as those proposed
by \citep{Bernal:2011qz,Barrientos:2016ipx}). One potential issue
with modified gravity is Soussa-Woodard's no-go theorem \citep{Soussa:2003sc,Soussa:2003mp},
which states that a metric-based, covariant gravitational theory exhibiting
MOND-like behavior cannot be stable. Consequently, there have been
attempts to obtain MOND-like behavior by abandoning covariance \citep{Milgrom:2019rtd}
and the equivalence principle \citep{Smolin:2017kkb}, particularly
in the deep-MOND regime, where the conflict between modified gravity
and these fundamental principles becomes more pronounced. In contrast,
Modified Inertial approaches can preserve the elegance of covariant
gravitational theories as much as possible, and the concept of acceleration,
which plays a foundational role in MOND, appears more naturally in
kinematics than in metric-based gravitational theories. Along this
line, Milgrom proposed an explanation for modified inertial behavior
based on the interaction between the Unruh temperature effect induced
by acceleration and the temperature effect of the background de Sitter
spacetime \citep{Milgrom:1998sy}. This idea has inspired numerous
similar attempts and improvements \citep{Pikhitsa:2010nd,Kiselev:2010xi,Klinkhamer:2011un,Ho:2011xc,Pazy:2011vx,Verlinde:2016toy,Smolin:2017kkb,Milgrom:2018bit,Alexander:2018lno}.
Both the deep-MOND limit and de Sitter geometry exhibit similar scale
invariance \citep{Milgrom:2016scu}, and the newly introduced critical
acceleration constant in MOND is of the same order of magnitude as
the cosmological constant in de Sitter geometry. These coincidences
have led to widespread speculation that the influence of the deSitter
cosmological background on locally accelerated systems may be the
cause of MOND-like modifications \citep{Smolin:2017kkb,Milgrom:2020cch}.
However, these speculations still lack a solid physical foundation
and mathematical proof. On one hand, the known effects of deSitter
background expansion on local gravitational systems like galaxies
seem negligible \citep{Cooperstock:1998ny}. Additionally, the static
solution of Einstein's gravity with a cosmological constant, the so-called
Schwarzschild-deSitter metric, fails to produce the $\Phi\sim\ln r$
gravitational potential and correct rotation curves required by MOND.
Furthermore, the Unruh effect arises from long-time uniform acceleration
in a flat background, which is quite different from the non-uniform
acceleration that occurs in real galactic acceleration systems. Moreover,
the question of how effects on the spectrum, such as those similar
to the Unruh effect, influence the classical equations of motion for
test particles has not yet been established on a firm physical basis.
In our view, a more complete theory addressing these issues must first
generalize the Unruh effect to non-flat, non-long-time non-uniform
acceleration scenarios, which is also an important independent physics
problem in its own right, even without the motivation of understanding
MOND. Second, it must explain how changes in the quantum spectrum
of test particles reflect their classical equations of motion. In
our opinion, although MOND still faces many unresolved issues, it
serves as an excellent phenomenological touchstone for testing many
new ideas. If, as many studies suggest \citep{Milgrom:1998sy,Verlinde:2016toy,Smolin:2017kkb,Milgrom:2020cch},
the fundamental theory behind MOND is a quantum gravity theory, then
MOND may hint at the most important features of this underlying quantum
gravity theory, which is a fascinating question indeed.

In our previous work \citep{Luo:2023eqf}, we employed the Gabor transform
method, a form of Fourier transform with a short-time window, to extract
spectral information from a monochromatic wave function undergoing
a local short-time acceleration within a brief time window. We demonstrated
that its spectrum undergoes a slight Gaussian broadening (an extra
broadening beyond the intrinsic standard quantum Gaussian broadening).
This second moment broadening of the spectrum is directly proportional
to the square of the instantaneous acceleration. In the limit of long-time
uniform acceleration, the spectrum reverts to the Planckian thermal
radiation spectrum predicted by the Unruh effect. Therefore, this
short-time acceleration-induced broadening effect can be regarded
as a local short-time generalization of the Unruh effect. This effect
can also be utilized to explain the influence of local non-uniform
acceleration on the kinematics of test particles therein, as well
as its impact on quantum spacetime itself through the quantum equivalence
principle \citep{2024arXiv240809630L} (where the universal part of
the second-order quantum fluctuations of test particles, serving as
quantum reference frames, measures the second-order quantum fluctuations
of quantum spacetime). This effect provides a more robust explanation
for the kinematic modifications to acceleration in a deSitter background
proposed by Milgrom \citep{Milgrom:1998sy}. In this paper, we will
derive the effect of spectral broadening due to the local short-time
acceleration using a more rigorous general coordinate transformation
method. Furthermore, we aim to leverage this effect to discuss, in
a more intuitive and enlightening manner, its role and relationship
within the quantum reference frame \citep{Luo2014The,Luo2015Dark,Luo:2015pca,Luo:2019iby,Luo:2021zpi,Luo:2022goc,Luo:2022statistics,Luo:2022ywl,2023AnPhy.45869452L,2024arXiv240809630L,Luo:2025cer,Luo:2025fdh}
and quantum gravity frameworks that we have established in the past.
We particularly emphasize that this effect is more naturally interpreted
as a modified kinematics or modified acceleration of test particles.
In our finalized theoretical formulation of quantum reference frames,
the form of gravity is strongly constrained by its fundamental principles,
with its infrared behavior closely resembling that of Einstein-Hilbert
gravity plus a cosmological constant (with major gravitational modifications
occurring in the ultraviolet regime). There is limited room for significant
modifications in the infrared behavior of gravity. While the infrared
gravitational behavior does deviate from Newtonian gravity in local
gravitational systems due to the cosmological constant, we recognize
that our theory cannot extend this deviation trend into the deep-MOND
regime \citep{Luo:2022ywl}, making it difficult to fully account
for the modified gravity behavior required by MOND. However, we realize
that the theoretical framework of quantum reference frames essentially
achieves quantum gravity by quantizing the concept of reference frames,
a modification of fundamental kinematics. One of the core concepts
in this theory, namely the universal second-moment quantum fluctuations
or broadening of the quantum frame field and the quantum equivalence
principle \citep{2024arXiv240809630L}, plays a crucial role in understanding
MOND. The behavior demanded by MOND can be naturally incorporated
into the quantum reference frame theory, which is one of the important
motivations for our current work. Additionally, the Modified Kinematics
or Modified Acceleration interpretation naturally avoids the issues
posed by Soussa-Woodard's no-go theorem, making it a highly promising
direction worthy of further exploration.

Another important motivation of the paper is, as part of a series
of previous works, continuously emphasizing the importance of the
second moment quantum fluctuations in quantum spacetime reference
frames, quantum gravity, and in understanding the cosmological observational
anomalies, which is currently an aspect that has not yet received
sufficient attention. We have discussed its origin in quantum spacetime
reference frames fields \citep{Luo2014The,Luo:2015pca}, its contradiction
and reconciliation with the equivalence principle \citep{2024arXiv240809630L},
the fundamental theory whereby second moments induce the Ricci flow
of spacetime \citep{Luo:2021zpi,Luo:2022goc,Luo:2022statistics,2023AnPhy.45869452L,Luo:2025cer},
how second moments naturally lead to a theory of quantum gravity coming
from the quantum theory of reference frame and the quantum equivalence
principle \citep{2024arXiv240809630L,Luo:2021zpi}, its role in establishing
an intrinsic geometric and relational quantum framework to describe
the relation/entanglement between a quantum system and quantum observer
or quantum reference system \citep{Luo:2025fdh}, and we have also
discussed how the second moment of quantum clock \citep{Luo2014The}
and quantum spacetime reference system \citep{Luo:2015pca,Luo:2021zpi}
lead to a resolution of the cosmological constant problem, as well
as how the second moment gives rise to the accelerated expansion of
the universe \citep{Luo2015Dark} and the radial acceleration discrepancies
\citep{Luo:2023eqf}, among other topics.

The structure of this paper is as follows: In Section II, we calculate
the correction to the action and the broadening effect on test particles
caused by local short-time acceleration. In Section III, we employ
a similar method to compute the correction to particle actions in
a deSitter background, as well as the equivalent comoving background
acceleration. Section IV discusses the scenario where both a deSitter
background and acceleration are present, and derives the acceleration
interpolation relation for MOND. Finally, in Section V, we provide
some discussions and a summary of our findings.

\section{Local Short-Time Acceleration correction to Inertial Particle Action}

We start from a general relativistic point particle action
\begin{equation}
S=m\int ds\sqrt{g_{\mu\nu}\frac{dx^{\mu}}{ds}\frac{dx^{\nu}}{ds}}\label{eq:general action}
\end{equation}
where $m$ is the mass, $x^{\mu}$, $(\mu=0,1,2,3)$ are the 4-coordinates
of the particle, $ds^{2}=g_{\mu\nu}dx^{\mu}dx^{\nu}$ is the proper
time, $g_{\mu\nu}$ is a general background metric where the particle
lives in.

If we consider that the particle is undergoing acceleration, a rigorous
approach to obtain the action for this accelerated particle is to
directly employ the metric $g_{\mu\nu}$ of the accelerated frame.
If the particle is non-relativistic, an alternative non-relativistic
approximation method is to use the flat metric $\eta_{\mu\nu}$ of
an inertial or flat coordinate system, but perform a coordinate transformation
on the particle's coordinates $x^{\mu}$ from this proper inertial
frame to the accelerated frame
\begin{equation}
x^{\mu}\rightarrow x^{\mu}+a^{\mu}\delta s^{2}\label{eq:coordinate transform}
\end{equation}
where $a^{\mu}(x)=\left(0,a^{i}(x)\right)$, $(i=1,2,3)$ is a non-relativistic
and non-uniform acceleration, $\delta s$ represents a very short
proper time interval during acceleration, satisfying
\begin{equation}
a^{\mu}\delta s\ll1
\end{equation}

Without loss of generality, we have assumed in the coordinate transformation
(\ref{eq:coordinate transform}) that the relative velocity of the
accelerated frame with respect to the proper inertial frame is zero.
If the relative velocity is not zero, the resulting effect is merely
that a boost introduces a Lorentz factor, which performs a Lorentz
transformation on the proper time $ds$. However, this is not the
focus of this paper, instead, our concern lies with the boost caused
by a local short-time acceleration.

Since that this acceleration $a(x)$ is short-time and occurs only
locally in space, we can consider that this coordinate transformation
can shift the particle from the proper inertial frame to a general
non-uniform and ``instantaneous'' accelerated frame with an ``instantaneous
acceleration'' (although, in the quantum spectral sense of the particle,
``instantaneous'' does not imply an infinitely short duration as
in classical physics, but rather at least a finite characteristic
time interval of the spectrum; thus, more rigorously, it should be
termed ``short-time''). Therefore, the treatment here does not have
to be confined, as in the derivation of the Unruh effect, to very
specific uniformly accelerated flat coordinate systems (Rindler coordinates),
but can instead handle a non-uniform ``instantaneous'' accelerated
frame.

Under the coordinate transformation we obtain an accelerated action
on the flat metric $\eta_{\mu\nu}$
\begin{align}
S_{acc} & =m\int ds\sqrt{\eta_{\mu\nu}\frac{d}{ds}\left(x^{\mu}+a^{\mu}\delta s^{2}\right)\frac{d}{ds}\left(x^{\nu}+a^{\nu}\delta s^{2}\right)}\nonumber \\
 & =m\int ds\sqrt{Z_{acc}\eta_{\mu\nu}\frac{dx^{\mu}}{ds}\frac{dx^{\nu}}{ds}}\label{eq:accelerated action}
\end{align}
To the order $O(\delta s^{2})$, we have
\begin{equation}
Z_{acc}\approx1+a^{2}\delta s^{2}\label{eq:Z_acc}
\end{equation}
where $a^{2}=\eta_{\mu\nu}a^{\mu}a^{\nu}$. And within the factor
$Z_{acc}$, the first-order term in $\delta x$, i.e. $2\eta_{\mu\nu}\left(a^{\mu}\delta x^{\nu}+a^{\nu}\delta x^{\mu}\right)$,
has been omitted. Because, first, the variation w.r.t. the first-order
$O(\delta x)$ or $O(\delta s)$ merely yields the standard classical
geodesic equation of motion for the test particle and is not important
in its quantization, so it does not produce the required MOND correction,
and second, it is also noted that $\eta_{\mu\nu}a^{\mu}\frac{dx^{\nu}}{ds}=\eta_{\mu\nu}\frac{du^{\mu}}{ds}u^{\nu}=\frac{d}{ds}\left(u^{\mu}u_{\mu}\right)=0$.
As we will see later in the paper, the correction needed for MOND
arises from the second-order term $O(\delta x^{2})$ or $O(\delta s^{2})$
upon quantization, i.e. $a^{2}\delta s^{2}$, and hence, MOND is considered
as a second-order effect.

$Z_{acc}$ can be regarded as a modification factor arising from boosting
from an inertial frame to a coordinate system with zero relative velocity
but undergoing local short-time acceleration. In the general case
of non-zero relative velocity, $Z_{acc}$ incorporates both the velocity-dependent
part of the Lorentz boost factor and the acceleration-dependent part.
Therefore, it can be viewed as a generalized Lorentz factor under
conditions of local short-time acceleration. In this sense, this treatment
can clearly be interpreted as modifying kinematics, since Lorentz
contraction in the context of special relativity is also understood
as a modification of kinematics, rather than as a mechanical contraction
of rulers due to some forces. $Z_{acc}\eta_{\mu\nu}$ can be considered
as an effective metric for this local short-time accelerated frame.
All effects of local short-time acceleration relative to the inertial
frame $\eta_{\mu\nu}$ are approximately encoded within this factor.

Next, let's examine the new effects brought about by the correction
terms in this factor. Given that $\left|a^{2}\delta s^{2}\right|\ll1$,
the deviation from the flat metric and the action in the inertial
frame caused by this acceleration is a small quantity, resulting in
only a minor correction to the particle's motion. If we consider $\omega$
as the Fourier mode or proper frequency/spectrum of the particle's
coordinate $x(s)$,
\begin{equation}
x(s)=\int_{-\infty}^{\infty}\frac{d\omega}{2\pi}\,e^{-i\omega s}\,\tilde{x}(\omega)
\end{equation}
and considering that the short acceleration proper time $\delta s$
is taken as the finite characteristic time interval of the spectrum,
with $\delta s^{2}\sim\frac{1}{\omega^{2}}$ (in the natural units
where $\hbar=1$ and $c=1$), the action of the accelerated particle
(\ref{eq:accelerated action}) can then be expressed in the Fourier
mode space as
\begin{align}
S_{acc} & =m\int\frac{d\omega}{2\pi}\sqrt{\eta_{\mu\nu}\left(\omega^{2}+a^{2}\right)\tilde{x}^{\mu}\tilde{x}^{\nu}}\nonumber \\
 & =m\int\frac{d\omega}{2\pi}\sqrt{\eta_{\mu\nu}\frac{\tilde{x}^{\mu}\tilde{x}^{\nu}}{2\sigma_{\tilde{x}}^{2}}}
\end{align}
in which the coefficient
\begin{equation}
\sigma_{\tilde{x}}^{2}=\frac{1}{2\left(\omega^{2}+a^{2}\right)}
\end{equation}
preceding this coordinate quadratic form can be regarded as an isotropic
second-order moment quantum fluctuation $\langle\tilde{x}^{2}\rangle$
of the quantized coordinate operator after the system undergoes quantization.

Correspondingly, the particle's spectrum exhibits second-order moment
quantum fluctuations or broadening $\sigma_{\omega}^{2}=\langle\omega^{2}\rangle=\langle\omega\rangle^{2}+a^{2}$,
where the first term represents the quantum uncertainty inherently
carried by the wave-like nature of the particle's spectrum, while
the second term 
\begin{equation}
\langle\delta\omega^{2}\rangle=\langle\omega^{2}\rangle-\langle\omega\rangle^{2}=a^{2}\label{eq:acc_broaden}
\end{equation}
is an extra second-order moment Gaussian broadening of the particle's
spectrum induced by acceleration (differs by a factor of 2, i.e.,
$\langle\delta\omega^{2}\rangle=(2a)^{2}$, compared with the derivation
via the Gabor transform method in \citep{Luo:2023eqf}, which arise
from some detail differences). Since this term is squared, both acceleration
and deceleration result in a broadening of the spectrum. Such second-order
moment fluctuations in the spectrum or coordinates represent quantum
fluctuations or uncertainties that is beyond the standard quantum
mechanical framework typically defined in inertial frames.

The term ``broadening'' used throughout the paper for both spectrum
and coordinates refers to general extra second-order moment fluctuations,
without specifically implying that the width is broadened or narrowed.
In fact, it depends on the signature convention, when $\eta_{\mu\nu}=(+,-,-,-)$,
$a^{2}=\eta_{\mu\nu}a^{\mu}a^{\nu}<0$, it is narrowed; alternative,
when $\eta_{\mu\nu}=(-,+,+,+)$, $a^{2}=\eta_{\mu\nu}a^{\mu}a^{\nu}>0$,
it is broadened. Throughout the paper, we take the latter for convenience
in which the spatial distance is positive-defined. Further more, due
to the complementary nature of second-order moment fluctuations in
coordinates and spectrum as dictated by the standard quantum mechanical
uncertainty principle (as the leading-order approximation), we observe
that under local short-time acceleration, the spectrum ``broadens''
while the coordinates ``narrow''.

It is noteworthy that this extra broadening is independent of the
particle's mass, energy, and mode, and instead universally reflects
the acceleration of the local spacetime or coordinate system itself.
On one hand, this arises as a consequence of modifying kinematics
(since kinematics pertains to the concepts of space and time, rather
than the particles themselves). On the other hand, it is also a requirement
of the quantum equivalence principle \citep{2024arXiv240809630L},
which states that the universal (mass-independent, non-Hamiltonian-dynamical
and independent to the detail properties of the particle) quantum
second-order moment fluctuations of a test particle can be abstracted
out to measure the quantum fluctuation properties of the background
spacetime, rather than being simply interpreted as the intrinsic quantum
properties of the test particle alone. This forms the physical foundation
of quantum reference frame theory.

Since the universal second-order moment broadening of coordinates
or spectra reflects the intrinsic quantum properties of spacetime
itself, these second-order moment fluctuations modify the distance
quadratic form in Riemannian spacetime geometry. Consequently, Riemannian
spacetime geometry no longer preserves the metric quadratic form (non-isometric)
at the quantum level. For this very reason, the Jacobian arising from
general coordinate transformations alters the functional integral
measure, leading to diffeomorphism anomalies in quantum spacetime
\citep{Luo:2021zpi,2024arXiv240809630L}. The change in the action
under accelerated transformations (relative to the action in an inertial
frame) can essentially be regarded as a consequence of quantum anomalies
induced by general coordinate transformations. However, the inertial
forces (gravitational forces) resulting from these anomalies have
already manifested such changes in the variation of the effective
classical action.

If the Lorentz factor generated by a velocity boost changes the mean
values (first-order moments) of spacetime coordinate $\langle x\rangle$,
then the effect of a local short-time accelerated boost will induce
an extra second moment broadening of coordinate $\langle\delta x^{2}\rangle$
(beyond the conventional intrinsic quantum second moment uncertainty).
In the dual frequency domain, if the velocity of a particle's wavefunction
can be measured through the Doppler shift (redshift or blueshift)
of the mean value of its spectral line $\langle\omega\rangle$, then
acceleration can be measured through the extra broadening of its spectral
line $\langle\delta\omega^{2}\rangle$. Although we know that there
are numerous contributions to spectral line broadening, such as lifetime
broadening, thermal broadening, and other dynamical (non-universal)
broadenings, if future experimental precision allows for the careful
subtraction of these non-universal broadenings, the remaining universal
broadening will reflect the intrinsic acceleration and the nature
of spacetime itself.

\section{deSitter Spacetime correction to Inertial Particle Action}

For the calculation of the action of a point particle in a deSitter
spacetime background, we can follow a method similar to that in the
previous section. To compare with the short-time acceleration effect
in the previous section, which was merely a small correction to the
flat metric, we assume here that the curvature radius of the deSitter
spacetime background is much larger than the particle's scale. Consequently,
the curvature of spacetime experienced by the particle also represents
only a small correction to the flat metric $\eta_{\mu\nu}$ (throughout
the paper we call the comoving (free-falling) observer in the deSitter
universe with the leading flat metric $\eta_{\mu\nu}$, an ``inertial
observer'', such as an observer in a center-of-mass frame of a galaxy
immersed in the deSitter universe). Therefore, we expand the metric
up to the second-order around the flat metric
\begin{equation}
g_{\mu\nu}=\eta_{\mu\nu}-\frac{1}{3}R_{\mu\alpha\nu\beta}\delta x^{\alpha}\delta x^{\beta}+...\label{eq:dS metric}
\end{equation}
where $\left|R_{\mu\alpha\nu\beta}\delta x^{\alpha}\delta x^{\beta}\right|\ll1$,
and the Riemannian curvature of the deSitter background can be written
by a positive cosmological constant $\Lambda$ as
\begin{equation}
R_{\mu\alpha\nu\beta}=\frac{\Lambda}{3}\left(g_{\mu\nu}g_{\alpha\beta}-g_{\mu\beta}g_{\alpha\nu}\right)\label{eq:Rieman curvature}
\end{equation}
Similar to the approach in the previous section, we also neglect the
first-order terms in $\delta x$, namely $\left(\eta_{\mu\sigma}\Gamma_{\nu\rho}^{\sigma}+\eta_{\nu\sigma}\Gamma_{\mu\rho}^{\sigma}\right)\delta x^{\rho}$,
because this term affects the standard classical geodesic equation
of motion for a test particle, which is unimportant in the quantization,
and can also be canceled by a proper choice of a frame (non-covariant).
The significant quantum-level corrections, as well as those required
by MOND, arise from the quadratic terms in $\delta x$.

Substituting the metric into the general action (\ref{eq:general action}),
and taking the isotropic quantum expectation $\langle\delta x^{\mu}\delta x^{\nu}\rangle\approx\frac{1}{4}\eta^{\mu\nu}\delta s^{2}$,
we obtain
\begin{align}
S_{dS} & =m\int ds\sqrt{\left(\eta_{\mu\nu}-\frac{1}{3}R_{\mu\alpha\nu\beta}\delta x^{\alpha}\delta x^{\beta}\right)\frac{dx^{\mu}}{ds}\frac{dx^{\nu}}{ds}}\nonumber \\
 & =m\int ds\sqrt{Z_{dS}\eta_{\mu\nu}\frac{dx^{\mu}}{ds}\frac{dx^{\nu}}{ds}}
\end{align}
in which the quantum renormalization factor in the deSitter background
is
\begin{equation}
Z_{dS}\approx1-\frac{\Lambda}{12}\delta s^{2}\label{eq:Z_dS}
\end{equation}
By comparing this factor with (\ref{eq:Z_acc}), we can observe that,
for a uniformly isotropic deSitter background, a uniformly isotropic
background acceleration
\begin{equation}
a_{bg}\equiv\sqrt{\frac{\Lambda}{12}}
\end{equation}
gives a positive correction to the factor and hence it cancel the
negative correction terms in $Z_{dS}$, resulting in a flat metric
(as indicated by the equivalence principle, which states that in this
comoving accelerated frame, the particle no longer experiences the
gravitational effect of the deSitter background). Therefore, a test
particle in a deSitter background will undergo geodesic accelerated
motion with a uniform and isotropic background acceleration $a_{bg}$.
Based on the current measured value of accelerated expansion $a_{bg}^{2}\approx-\frac{1}{4}q_{0}H_{0}^{2}$,
we have $a_{bg}\approx5\times10^{-10}m/s^{2}$, where $q_{0}\approx-0.64$
is the cosmic deceleration parameter (the negative sign indicates
that the expansion is accelerating), and $H_{0}$ is the currently
measured Hubble constant.

Based on the considerations similar to those in the previous section,
a test particle undergoing comoving acceleration in a deSitter background
will produce an extra broadening of its spectrum $a_{bg}^{2}=\frac{1}{12}\Lambda\approx-\frac{1}{4}q_{0}H_{0}^{2}$
relative to an earth observer who considers themselves is in a flat
background and that mutually accelerated recessing away from it. This
extra broadening is indistinguishable from the universal extra broadening
(\ref{eq:acc_broaden}) caused by general acceleration. In fact, observational
results from the late-epoch universe dominated by a positive cosmological
constant (or dark energy) and experiencing accelerated expansion could
entirely stem from the correction to the distance-redshift relation
at the second-order in redshift, introduced by an extra second-order
moment broadening of spectral lines from Type-Ia supernovae (or other
test particles undergoing comoving expansion with high redshift and
negligible proper motion) (see previous papers \citep{Luo2015Dark,Luo:2021zpi,Luo:2023eqf,2024arXiv240809630L}):
If the luminosity distance ($d_{L}$) - redshift ($z$) relation is
expanded up to the quadratic order in redshift, it takes the form
\begin{equation}
\langle d_{L}\rangle=\frac{1}{H_{0}}\left[\langle z\rangle+\frac{1}{2}\langle z^{2}\rangle+...\right]=\frac{1}{H_{0}}\left[\langle z\rangle+\frac{1}{2}\left(1-q_{0}\right)\langle z\rangle^{2}+...\right]\label{eq:rdistance-redshift}
\end{equation}
where the quadratic term in redshift $\langle z^{2}\rangle=\langle z\rangle^{2}+\langle\delta z^{2}\rangle=\left(1-q_{0}\right)\langle z\rangle^{2}=Z_{r}\langle z\rangle^{2}$
receives an extra correction from the redshift broadening $\langle\delta z^{2}\rangle$.
This extra broadening of spectral lines reflects the accelerated expansion
of the universe, just as the universal cosmological redshift of spectral
lines indicates the Hubble expansion. In other words, the first moment
distance and redshift measurements give rise to the first moment of
Hubble constant, remains unaffected by the second moment of the redshift;
while the Hubble expansion rate as the first derivative is modified
by the positive second moment, so the expansion is regarded accelerating.
The (nearly) uniform and isotropic nature of dark energy implies that
this spectral line broadening is also nearly universal, independent
of the spectral line's energy. Consequently, the universal broadening
of spectral lines measures the accelerated expansion of the space
itself, as required by the quantum equivalence principle: all spectral
lines of different energies universally ``free fall'' (accelerate
away from each other) along geodesics with acceleration $a_{bg}$
in the cosmic spacetime background. From the perspective of the quantum
equivalence principle, the broadened spectral lines of Type-Ia supernovae
reflect a real expansion acceleration of the spatial grid, and the
two are equivalent and indistinguishable.

\section{Modified Acceleration upon a deSitter Background}

With the results of $Z_{acc}$ and $Z_{dS}$ derived in the previous
two sections, the action for a test particle undergoing local short-time
acceleration in a deSitter spacetime background can be written as
\begin{equation}
S_{acc+dS}=m\int ds\sqrt{Z_{eff}\,\eta_{\mu\nu}\frac{dx^{\mu}}{ds}\frac{dx^{\nu}}{ds}}\label{eq:particle action}
\end{equation}
where the effective modification factor is 
\begin{equation}
Z_{eff}\approx1+\left(a^{2}-\frac{\Lambda}{12}\right)\delta s^{2}\equiv1+a_{eff}^{2}\delta s^{2}\label{eq:Z_eff}
\end{equation}
and
\begin{equation}
a_{eff}^{2}=a_{T}^{2}-\frac{\Lambda}{12}\label{eq:acc+dS}
\end{equation}
Here, $a_{T}$ represents the Total acceleration of the test particle
relative to an ``inertial observer'' in the center-of-mass frame
of the galaxy immersed in the deSitter spacetime. Since the deSitter
background is equivalent to a background acceleration $a_{bg}\equiv\sqrt{\frac{\Lambda}{12}}$,
the total acceleration $a_{T}$ includes contributions from both the
particle's proper motion acceleration $a_{pp}$ relative to the spacetime
grid and the background acceleration $a_{bg}$ due to the accelerated
expansion of the spacetime grid itself. The vector summation of these
two acceleration contributions arises from the first-order terms of
$\delta x$ in $Z_{acc}$ and $Z_{dS}$, namely the combination of
$2\eta_{\mu\nu}\left(a^{\mu}\delta x^{\nu}+a^{\nu}\delta x^{\mu}\right)$
and $\eta^{\mu\nu}\left(\eta_{\mu\sigma}\Gamma_{\nu\rho}^{\sigma}+\eta_{\nu\sigma}\Gamma_{\mu\rho}^{\sigma}\right)\delta x^{\rho}$.
This combination affects only the mean value of acceleration (first-order
moment) and the classical geodesic equation of motion for the test
particle, without influencing the second-order moment of the particle's
spectrum (or coordinates), so although it can not produce the MOND
phenomenon, it is still an unavoided part \citep{Cooperstock:1998ny}.
Classically, because the acceleration due to background expansion
is uniform and isotropic, outwardly divergent at any spatial point,
the proper motion acceleration of the particle in any direction will
be superimposed upon this background acceleration in the same direction,
so
\begin{equation}
a_{T}^{\mu}=a_{pp}^{\mu}+a_{bg}^{\mu}\label{eq:classical composition}
\end{equation}
which is a component modification of a proper acceleration $a_{pp}$
by the background acceleration $a_{bg}$ at the first-moment level.

To produce MOND phenomenon, one need to consider the $Z_{eff}$ factor.
We can see that the factor $Z_{eff}$ together with (\ref{eq:acc+dS})
formally appears to modify the classical inertia of the free particle,
$m\rightarrow m_{eff}=m\sqrt{Z_{eff}(a,\omega)}$ as an inertia-modifying
interpolating function of MOND. As the modified inertial interpretation
of MOND, this modifying factor of course change the classical kinematics
and geodesic equation of the particle in the sense that the gradient
of the $Z_{eff}$ factor introduces an extra effective ``non-inertial
force'' into the geodesic equation of the particle. However, we argue
that this modified-inertial interpretation will break the covariance
and the equivalence principle. Because, as the shortcomings of MOND,
it makes the theory valid only for absolute preferred inertial frames,
and the inertial mass also does not equivalent to the gravitational
mass. Alternatively, if one notes that $Z_{eff}$ is merely a covariantly
changed factor of the action under a general coordinate transformations,
one can regard it does not modify the classical inertial and the classical
geodesic equation of motion (by the extra ``force''), since the
extra ``force'' can be absorbed into the coordinate transformed
new metric $Z_{eff}\eta_{\mu\nu}$ to make the geodesic equation formally
unchanged (while Levi-Civita connections changed corresponding to
the new metric), in this case, the covariance and equivalence principle
keep valid. Therefore, in the following paper, we will give an equivalent
interpretation of the modified particle action (\ref{eq:particle action})
that the real significance of this $Z_{eff}$ factor lies at the quantum
level, where it gives rise to a second moment correction to the particle
spectrum and the squared acceleration.

The eq.(\ref{eq:coordinate transform}) under the factor $Z_{eff}$
effective gives $x^{\mu}\rightarrow x^{\mu}+a_{eff}^{\mu}\delta s^{2}$,
where the effective acceleration $a_{eff}$ modifies the kinematic
acceleration $a$ in (\ref{eq:Z_acc}). $a_{eff}$ as a modified Levi-Civita
connection, can be derived by the gradient of the modified metric
$Z_{eff}\eta_{\mu\nu}$ in the particle action (\ref{eq:particle action}),
i.e. 
\begin{equation}
a_{eff}^{\mu}=\frac{1}{2}\frac{\partial Z_{eff}}{\partial x_{\mu}}\label{eq:a_eff}
\end{equation}
So the second moment quantum effect in $Z_{eff}$ effectively modifies
the classical acceleration vector. This phenomenon, where the second
moment effect effectively influences the mean of the first-order derivative,
also appears in the distance-redshift relation (\ref{eq:rdistance-redshift}),
that is, the second-order correction in the distance-redshift relation
will lead to an apparent enhance in the first-order derivative (deviating
from the linear Hubble's law), even if the central value (first moment)
of the Doppler redshift of the spectral line is not altered by the
second moment broadening. Therefore, the origin of acceleration anomalies
lies in the importance of the intrinsic second moment quantum fluctuations
in the deSitter universe, which gives rise to the fact that the mean
value of the gradient (the effective acceleration (\ref{eq:a_eff}))
does not equal to the gradient of the mean value (the first moment
acceleration given by the unmodified Newtonian gravity).

By taking the accelerating coordinate interval $\delta x^{\mu}=a_{eff}^{\mu}\delta s^{2}$
in (\ref{eq:a_eff}), it consistently gets the effective acceleration
square $a_{eff}^{2}=\frac{1}{2}a_{eff}^{\mu}\frac{\partial Z_{eff}}{\partial x^{\mu}}=\frac{1}{2}\frac{\partial^{2}Z_{eff}}{\partial s^{2}}$
in the expansion of $Z_{eff}$ at the second-order. In general, the
second moment quantum fluctuation manifests its correction effects
by modifying any physical quantity that is expanded up to the second-order
of the spectrum (or coordinates $x$), which follows the same principle
to the squared-order correction in the distance-redshift relation
(\ref{eq:rdistance-redshift}). 

Although the spectral broadening effect of $Z_{eff}$ and its gradient,
the directional effective acceleration vector $a_{eff}^{\mu}$, are
difficult to measure in most cosmological observations, there exist
some physical quantities that depend solely on the square or magnitude
of acceleration $a_{eff}^{2}$ at the second-order of $Z_{eff}$.
For instance, in the context of virial equilibrium (which is generally
the common method employed in cosmological observations to infer acceleration),
what matters is merely the relationship between (temporally and directionally)
averaged values of kinetic and potential energies of a test particle,
independent of the direction of force or acceleration. In this scenario,
the second-order effect corrected square or magnitude of the effective
acceleration, rather than directional components, plays a pivotal
role and directly manifests its real physical effects. Such a direction-less
property also holds true in the measurement of cosmic background acceleration.
The cosmic expansion acceleration is measured via the extra distortion
derived by expanding the distance-redshift relation (\ref{eq:rdistance-redshift})
in the squared term. Currently, it can only be measured that the background
acceleration is either in the same direction as the Hubble expansion
velocity (spectral line broadening, $q_{0}<0$) or in the opposite
direction (spectral line narrowing, $q_{0}>0$). No other non-collinear
angle between the acceleration and the expansion velocity can be obtained
through the single redshift squared term (unless the squared term
is not isotropic and uniform, unlike the linear term (the Hubble expansion
rate), but current observations have little evidence for such an anisotropy).
So in general, an isotropic and uniform (temporal) spectral broadening
(\ref{eq:acc_broaden}) does not contain information about the direction
of the acceleration, but it is enough to imply many acceleration anomalies
in cosmology, including the accelerated expansion of the universe
based on the distance-redshift relation and the acceleration discrepancies/rotation
curve anomaly in galaxies based on the virial equilibrium.

\section{Quantum Equivalence Principle}

The fact that the effective acceleration (\ref{eq:acc+dS}) is equivalent
and indistinguishable to a physical and real acceleration requires
a quantum equivalence principle \citep{2024arXiv240809630L} as its
physical foundation. The principle claims that the universal part
of the second moment fluctuation (universal means that e.g. the quantum
fluctuation (\ref{eq:acc_broaden}) is independent to the detail properties
of the particle but only its acceleration) in the test particle's
spectrum (or coordinates) is indistinguishable from the square/magnitude
of acceleration or curvature/gravity, just as relativity requires
a classical equivalence principle to ensure that relativistic metric
effects (such as Lorentz contraction or time dilation) are not merely
apparent phenomena concerning just specific rulers or clocks, but
are universal real spacetime effects, with both being indistinguishable.
The classical equivalence principle only involves equivalence at the
level of the mean values (first-order moments), whereas the quantum
equivalence principle extends this equivalence to the second moments
(and even higher-order moments) universally. Thus, a quantum equivalence
principle ensures that, if one measures the physical quantity that
depends on the square/magnitude of the acceleration rather than its
components or directions, such as the distance-redshift relation (\ref{eq:rdistance-redshift})
or the velocity dispersion (which is governed by the square/magnitude
of acceleration via the virial equilibrium, see (\ref{eq:virial})),
the test particle appears to experience an effective acceleration
with the second-order moment fluctuation correction as if it is a
real acceleration. 

The relation (\ref{eq:Z_eff}) and (\ref{eq:acc+dS}) indeed embodies
the quantum equivalence principle: the acceleration and the curvature
(or equivalently, the cosmological constant) are at the same order
in the $Z_{eff}$ factor, meaning they are on an equal footing and
indistinguishable at the quantum level. Both of them broaden the spectrum
(though the microscopic broadening of spectrum is far too small to
be directly modify the macroscopic velocity dispersion from the Doppler
broadening), however, in this manner, the cosmological constant enters
as an acceleration, which modifies $a_{T}^{2}$ and gives rise to
a physical effective acceleration $a_{eff}^{2}$. At the quantum level,
acceleration, gravity (curvature), and spectral broadening are equivalent:
the classical equivalence principle establishes the equivalence of
the first two (at the first-moment level and each components level),
while \citep{2024arXiv240809630L} proposes to extend this equivalence
to include the spectral broadening (at the second-moment and square/magnitude
level), naming it the generalized quantum equivalence principle. The
universal (independent to the spectral line energies and spatial directions,
etc) broadening correction at the second-order expansion of the distance-redshift
relation (\ref{eq:rdistance-redshift}) is also shown as an verification
of such an equivalence between the spectral broadening and the accelerated
expansion of the space grid. The Unruh effect, which describes the
full broadening of the spectrum into a blackbody spectrum when one
transits from the short-time acceleration to a long-time limit acceleration,
is also regarded in extensive literature as another example of this
quantum equivalence between the first two and the third one (the relation
between the short-time slightly Gaussian broadening and the long-time
Unruh type full blackbody broadening can be found in \citep{Luo:2023eqf}).

Although the squared acceleration term in the $Z_{eff}$ factor is
too small to be directly measured via spectral broadening, it still
can be realized alternatively. There do exist several measurement
methods in cosmological observations that only involve the square
or magnitude of a galaxian acceleration, without any reference to
its direction, for instance, through a directionally averaged and
static virial equilibrium relation. We consider a simple virial equilibrium
astronomical system surrounding a point-like baryonic gravitational
source, such as an isolated galaxy immersed in the accelerated expanding
or the deSitter universe. Both the quantum equivalence principle and
the acceleration composition effect (\ref{eq:acc+dS}) jointly gives
the relation (the inertial mass and gravitational mass are assumed
equivalent, and both are not modified)
\begin{equation}
\langle\sigma_{\phi}^{2}\rangle+\langle\sigma_{\theta}^{2}\rangle+\langle\sigma_{r}^{2}\rangle=\langle\boldsymbol{r}\mathbf{\cdot}\boldsymbol{a}_{eff}\rangle_{dir}=r\sqrt{\overline{\cos^{2}\alpha}\langle a_{eff}^{2}\rangle}=\beta r\sqrt{\langle a_{eff}^{2}\rangle}\label{eq:virial}
\end{equation}
in which $\beta^{2}$ is a directionally averaged factor, formally
$\beta^{2}=\overline{\cos^{2}\alpha}$ ($\alpha$ the angle between
$a_{eff}$ and $r$), and if one consider the simple case that the
gravitational source is approximately point-like, without any prior
knowledge of the acceleration direction, we can assume $\beta^{2}=\overline{\cos^{2}\alpha}\sim\frac{1}{3}$,
if the averaged potential energy in virial equilibrium is considered
averaging over all direction in 3-space. It is because of the directionally
averaged $\beta^{2}$ factor, the virial equilibrium relation can
be considered only depend on the square or magnitude of the effective
acceleration, rather than its direction. So $\langle\boldsymbol{r}\mathbf{\cdot}\boldsymbol{a}_{eff}\rangle_{dir}$
denotes the directionally averaged virial potential energy given by
the magnitude of the effective acceleration (\ref{eq:acc+dS}) for
those almost point-like gravitational source. 

It is also worth arguing that, we regards the virial potential energy
given by the effective acceleration as a real virial potential energy,
it is manifested by the fact that the effective acceleration is nothing
but the (negative) gradient of a static gravitational potential encoded
in the effective metric $Z_{eff}g_{\mu\nu}$ (\ref{eq:a_eff}). Fundamentally,
its realness is based on the quantum equivalence principle. We consider
the principle, as a fundamental assumption for the quantum nature
of spacetime, cannot be derived from any deeper microscopic mechanisms,
that is to say, this paper neither attempt to prove how exactly the
real virial potential energy should be derived from the potential
energy of the known Newtonian acceleration $a_{N}$ plus some emergent
energy or effective pressure coming from the second moment fluctuations
by some deeper microscopic mechanisms, nor to establish a new macroscopic
virial equilibrium relation modified by some microscopic mechanism
of the second moment fluctuations. Of course these approaches are
viable, yet it is not the path we adopt. Instead, we take the quantum
equivalence principle as our fundamental postulate and starting point,
as it can provide a brand-new foundation for the quantum reference
frame and quantum gravity theory. In this manner, we directly regard
the virial potential energy $\langle\boldsymbol{r}\mathbf{\cdot}\boldsymbol{a}_{eff}\rangle_{dir}$
given by the effective acceleration as the real virial potential energy,
and those ``microscopic mechanism'' can be effectively emerged from
the quantum equivalence principle (e.g. the microscopic entropy of
quantum spacetime \citep{Luo:2022statistics,Luo:2025cer}), analogous
to directly regard the effective inertial mass (includes quantum self-energy
corrections) as the real gravitational mass, and the microscopic origin
of the gravitational mass corrections can be derived based on the
equivalence of these two kinds of mass. The quantum equivalence principle
governs all gravitational and spacetime effects related to the universal
second moment quantum fluctuations, and it can only be verified or
falsified through experimental measurements (more independent tests
and evidences of the quantum equivalence principle see \citep{2024arXiv240809630L}).
Although the short-time acceleration induced spectral broadening effect
provide a microscopic mechanism to make the cosmological constant
enter as an acceleration, it is a verification of the quantum equivalence
principle, rather than a microscopic foundation of the principle. 

Just like in the classical equivalence principle, we do not need to
resort to deeper classical mechanical principles to prove why, for
instance, locally the work done by acceleration (or related inertial
force) is identical to the work done by gravity. The classical equivalence
principle is a fundamental assumption or starting point, which allows
us to establish a geometric gravity theory upon the general coordinate
transformation or spacetime metric distortion induced by acceleration.
Similarly, the quantum equivalence principle is also a fundamental
assumption (without deeper microscopic mechanism), based on the principle,
we can further construct a geometric quantum gravity that incorporates
second moment quantum fluctuations upon the universal second moment
quantum fluctuations of the material coordinate system or spectral
lines as the measures of the quantum spacetime. The spacetime effective
distance quadratic form (effective metric), effective curvature, effective
acceleration, etc., modified by these geometric and universal second
moments all behave as if they were real physical quantities.

\section{Velocity Dispersion}

In the scenario of a spiral galaxy, $\langle\sigma_{r}^{2}\rangle$
and $\langle\sigma_{\theta}^{2}\rangle$ in (\ref{eq:virial}) as
the static radial and declinational velocity dispersion are considered
small. And $\langle\sigma_{\phi}^{2}(r)\rangle$ in (\ref{eq:virial})
is the static tangential velocity dispersion of the disk spiral galaxy
(the azimuthal direction) of the test particle at radius $r$ measured
through the Doppler broadening of spectral lines with some tilt projections
(approximately half the amplitude of the line width since the line
width includes the motions toward and away from an ``inertial observer''). 

In the scenario of disk spiral galaxy where $\langle\sigma_{r}^{2}\rangle$
and $\langle\sigma_{\theta}^{2}\rangle$ can be neglected, we can
simply attribute the squared effective acceleration $\langle a_{eff}^{2}\rangle$,
which is subject to correction (\ref{eq:acc+dS}), to the measured
tangential velocity dispersion $\langle\sigma_{\phi}^{2}\rangle$
by the virial equilibrium. We get the relation from (\ref{eq:virial})
\begin{equation}
\beta^{2}\langle a_{eff}^{2}\rangle\approx\frac{\langle\sigma_{\phi}^{4}(r)\rangle}{r^{2}}\label{eq:virial-2}
\end{equation}

As a consequence, by using (\ref{eq:acc+dS}), (\ref{eq:classical composition})
and (\ref{eq:virial-2}), the acceleration interpolation function
is
\begin{equation}
\beta^{-1}\frac{\langle\sigma_{\phi}^{2}(r)\rangle}{r}\approx\sqrt{\langle a_{eff}^{2}\rangle}=\sqrt{\left(a_{N}+a_{bg}\right)^{2}-a_{bg}^{2}}\label{eq:acc_composition}
\end{equation}
The acceleration composition relation is a universal (independent
of the specific stellar system) acceleration correction for an isolated
many-body virial equilibrium gravitational system, such as the rotation-supported
spiral galaxy isolately embedded in a deSitter cosmic background. 

When $a_{N}\gg a_{bg}$, this formula asymptotically recovers standard
Newtonian behavior; when $a_{N}$ is much smaller compared to $a_{bg}$
(in the deep-MOND regime), such as at the edges and beyond of isolated
spiral galaxies, this formula yields the required MOND dynamics
\begin{equation}
a_{eff}\approx\sqrt{2a_{N}a_{bg}},\qquad(a_{N}\ll a_{bg})
\end{equation}

It is worth stressing again that the absolute magnitude of this spectral
broadening effect by acceleration is extremely small and almost unmeasurable,
so the effect that accelerates the broadening of spectral lines does
not influence the measurements by directly altering the measured spectral
line width, since the acceleration broadening is far smaller than
the Doppler broadening of the spectral lines used to measure velocity
dispersion. The acceleration-broadened spectrum, which has an equivalent
spectral broadening effect to the background cosmological constant
or $a_{bg}$, contributes via eq.(\ref{eq:acc+dS}) to modify the
squared acceleration, thereby correcting the measured velocity dispersion
through the virial equilibrium. The eq.(\ref{eq:acc_composition})
reflects the competitive relationship between the two accelerations
that simultaneously contribute to the broadening: one is the broadening
from the classical Newtonian acceleration, and the other is the broadening
from the deSitter background acceleration. This deviation only becomes
significant when the Newtonian acceleration is small enough to be
comparable to the background acceleration. The relative comparison
between two effects that are both tiny in absolute magnitude can nevertheless
produce a pronounced observable effect.

If we substitute $a_{N}=\frac{GM}{r^{2}}$ (where $G$ is the Newtonian
constant and $M$ is the mass of baryonic matter of the galaxy), then
it yields the baryonic Tully-Fisher relation for spiral galaxies
\begin{equation}
v_{f}^{4}\approx2\beta^{2}GMa_{bg}
\end{equation}
in which $v_{f}^{2}=\lim_{r\rightarrow\infty}\langle\sigma_{\phi}^{2}(r)\rangle$
is the asymptotic rotational velocity at large radii. Taking $a_{bg}\equiv\sqrt{\frac{\Lambda}{12}}\approx2.5\times10^{-10}m/s^{2}$,
and $\beta^{2}=\overline{\cos^{2}\alpha}\sim\frac{1}{3}$ for the
approximate point-like gravitational source distribution, then the
coefficient $2\beta^{2}a_{bg}\approx1.5\times10^{-10}m/s^{2}$ is
slightly higher but very close to the fitting value $a_{0}\approx1.2\times10^{-10}m/s^{2}$
obtained from the observed Tully-Fisher relation $v_{f}^{4}=GMa_{0}$
for spiral galaxies. We argue that this tension does not fail the
idea, if the present effect is real, it has already provided a leading
scaling behavior to the Tully-Fisher relation, and there are still
several secondary contributions that may correct the calculation,
for example, the non-point-like gravitational source of a galaxy may
contribute a few percent to the potential term $\frac{GM}{r}$, and
the other possible renormalization effect or scale dependence effect.
In fact, the fitting of $a_{0}$ in the non-point-like and larger
scale objects, such as the galaxy clusters/groups is slightly larger
than the isolated spiral galaxies. This may due to the shape correction,
or the scale dependent effect, or the external field effect (the external
field effect is also exist in the framework but not exactly the same
with the one in MOND, since it is sensitive to the exact specific
interpolating function), etc. The exact origins of the tension are
still open. 

Unlike spiral galaxies, in elliptic galaxies, test particles with
velocity dispersion components in all directions exist. In this case,
$\langle a_{eff}^{2}\rangle$ cannot be solely related to $\langle\sigma_{\phi}^{2}\rangle$,
and the velocity dispersion in other directions are non-zero in eq.(\ref{eq:virial}).
However, if we consider that a random motion system like an elliptical
galaxy is approximately spherically symmetric and isotropic, meaning
that, statistically, the velocity dispersions (i.e. energies) of the
three components are evenly distributed, with the equipartition energy
reflected by the uniform projected dispersion velocity $\langle\sigma^{2}\rangle$
(which can be measured through Doppler broadening of spectral lines),
then $\langle a_{eff}^{2}\rangle$ can be related to the square of
the velocity dispersion $\langle\sigma^{4}\rangle$ by the same consideration
of the virial equilibrium. Therefore, we can qualitatively derive
a Faber-Jackson relation for elliptical galaxies, $\langle\sigma^{4}\rangle\sim GMa_{bg}$,
up to a numerical coefficient differing on both sides of the relation
depending on the specific profile of the galaxy, and for more general
pressure-supported systems, the relation is manifested as the general
mass-velocity-dispersion ($M-\sigma$) relation.

For the effects like gravitational lensing, it is clear that the $\sqrt{Z_{eff}}$
factor in the relativistic particle action (\ref{eq:particle action})
plays the role of an effective refractive index of light. The refractive
index change from the intergalactic space $\left.\sqrt{Z_{eff}}\right|_{a_{N}=0}=1$
to that within the galaxy $\left.\sqrt{Z_{eff}}\right|_{a_{N}\neq0}>1$
gives rise to the deflection of light. Therefore, the present effect
will not only modify the rotational velocity dispersion, but also
the gravitational lensing as conventional dark matter theory. The
details of the gravitational lensing produced by the present effect
is leaving for our future study. This is actually a manifestation
of the quantum equivalence principle: the correction this effect introduces
to the particle action (\ref{eq:Z_eff}) is independent of the particle's
mass, applying equally to massive particles in gravitationally bound
states and massless unbound photons, the correction is universal,
so it can be attributed to the correction of the spacetime metric
or the effective potential. More precisely, considering a relativistic
particle (\ref{eq:particle action}) in a static metric $g_{rr}=1-2\Phi_{N}$
with a Newtonian potential $\Phi_{N}$, the modified-kinematics $a_{pp}^{i}\rightarrow a_{eff}^{i}=\frac{1}{2}\nabla_{i}Z_{eff}$
in the flat metric can be equivalently interpreted as the modified-gravity
$g_{\mu\nu}\rightarrow Z_{eff}g_{\mu\nu}$ or $\Phi_{N}\rightarrow Z_{eff}\Phi_{N}$,
which leads to the conventional gravitational lensing. The quantum
equivalence principle ensures a symmetry between the modified-kinematics/acceleration
and the modified-gravity, it is completely different from the conventional
interpretation of MOND. In essential, the universal second moment
correction not only modifies kinematics/acceleration of test particles,
but also modifies spacetime and hence modifies gravity.

\section{Discussions and Conclusions}

(1) This paper interprets the acceleration interpolation relation
proposed by MOND, the relation is derived from the correction to the
action of a free particle on a flat background due to local short-time
non-uniform acceleration (and deSitter background), without the need
for artificial introduction. This interpolation relation arises from
a coordinate transformation, akin to a Lorentz boost factor, when
transitioning the free particle from a flat coordinate system to a
local short-time accelerated coordinate system (or a deSitter coordinate
system). Consequently, this quantum non-inertial effect modifies the
kinematics or acceleration of a test particle without modifying the
gravity theory, despite the interpretations of the modified kinematics/acceleration
and modified gravity are essentially equivalent in the sense of the
quantum equivalence principle.

(2) The effects induced by acceleration can be interpreted as an extra
second-order moment broadening of the spectrum (or coordinates) of
a test particle. For a test particle accelerating in a galaxy immersed
in the deSitter background, the effective acceleration (w.r.t. an
``inertial observer'' in the center-of-mass frame of the galaxy)
arising from the composition of its proper motion acceleration and
the background cosmological constant is considered physically coming
from the effective broadening of the spectrum is a composition of
the particle's proper motion acceleration broadening and the background
spectral broadening. In this manner, the cosmological constant enters
the kinematics as an acceleration. Beside the effect, the quantum
equivalence principle is another basis for MOND. That is, the universal
part of the second-order moment fluctuations in the spectrum (or coordinates)
of a test particle is indistinguishable from the acceleration of its
comoving coordinate system/spacetime grid, or from inertial/gravitational
forces or spacetime curvature, just as relativity requires an equivalence
principle to ensure that relativistic metric effects (such as Lorentz
contraction, time dilation, and spacetime curvature) are not merely
apparent phenomena but are real spacetime effects, with both being
indistinguishable. A quantum equivalence principle ensures that, for
a test particle accelerating in a deSitter background, the effective
acceleration resulting from the extra universal second-order moment
broadening of its spectrum (or coordinates) is a real acceleration.
Its physical realness is manifested in that the universal broadening
of spectral lines from Type-Ia supernovae reflects the real accelerated
expansion of the space background (\ref{eq:rdistance-redshift}) where
the proper motion accelerations are small, and also manifested finally
in that the background acceleration corrected squared effective acceleration
modifies the velocity dispersion of galaxies via the virial equilibrium
where the proper motion accelerations also contributes to the effective
acceleration, see (\ref{eq:classical composition}) and (\ref{eq:acc+dS}).
Both the acceleration induced spectral broadening effect and the quantum
equivalence principle jointly give rise to the interpretation of MOND.

(3) Strictly speaking, the term ``broadened'' refers to both spectra
and coordinates denotes general extra second-order moment fluctuations
or distortion, without specifying whether the width necessarily broadens
or narrows. In fact, the spectra is ``broadened'' or ``narrowed''
depends on the signature convention, when $\eta_{\mu\nu}=(+,-,-,-)$,
$a^{2}=\eta_{\mu\nu}a^{\mu}a^{\nu}<0$, it is narrowed in (\ref{eq:Z_acc});
alternative, when $\eta_{\mu\nu}=(-,+,+,+)$, $a^{2}=\eta_{\mu\nu}a^{\mu}a^{\nu}>0$,
it is broadened in (\ref{eq:Z_acc}). Throughout the paper, we use
the term ``broadened'' is just by taking the latter for convenience
in which the spatial distance is positive-defined. Due to the standard
quantum mechanical uncertainty principle (at least as the leading-order
approximation), the second-order moment fluctuations of coordinates
and spectra are complementary: the spectrum broadens while the coordinate
narrows.

(4) The paper has derived the broadening of the (temporal) spectrum
$g_{\mu\nu}\rightarrow Z_{eff}g_{\mu\nu}$ and the modification of
the squared acceleration, and has not addressed anisotropic (spatial)
momentum broadening that involve the directionality, it is because
the simple and intuitive effect is enough for a conceptual understanding
of rotation curve anomalies in many cosmological observation based
on the averaged and static virial equilibrium (see below) and the
accelerated expansion of the universe based on the distance-redshift
relation (\ref{eq:rdistance-redshift}), a full anisotropic off-diagonal
broadening effect $g_{\mu\nu}\rightarrow Z_{\mu}^{\alpha}Z_{\nu}^{\beta}g_{\alpha\beta}\approx g_{\mu\nu}+a_{\mu}a_{\nu}\delta s^{2}$
could be studied in a more general and complex case. And at the quantum
level, due to the quantum equivalence principle, the universal second
moment correction of the test particle (serving as a quantum reference
system) will in turn affect the spacetime metric itself driven the
Ricci curvature ($\delta g_{\mu\nu}\approx a_{\mu}a_{\nu}\delta s^{2}\propto R_{\mu\nu}$),
called Ricci flow, in other words, a full theory of the second-moment
broadening tensor is the Ricci flow theory (see e.g. \citep{2024arXiv240809630L,Luo:2025cer}),
which causes the spacetime metric to irreversibly evolve toward a
maximum-entropy and maximum-symmetry soliton configuration, such as
the deSitter configuration, which is beyond the scope of the paper. 

(5) Based on the quantum equivalence principle, the effect of such
acceleration-induced broadening of spectral lines (or coordinates)
is not merely the correction of spectrum itself, but also the universal
correction of the metric and curvature, which can be treated more
rigorously and generally (in non-uniform and anisotropic scenarios)
within the frameworks of quantum reference frames and quantum gravity
theories \citep{Luo2014The,Luo2015Dark,Luo:2015pca,Luo:2019iby,Luo:2021zpi,Luo:2022goc,Luo:2022statistics,Luo:2022ywl,2023AnPhy.45869452L,2024arXiv240809630L,Luo:2025cer,Luo:2025fdh}.
These approaches involve the Ricci flow of quantum spacetime (a renormalization
group flow of the metric tensor in quantum spacetime) or the broadening
of heat kernels associated with conjugate heat equations coupled to
the spacetime Ricci flow. Therefore, the isotropic broadening of spectra
(or coordinates) described in this paper serves as an intuitive and
useful concept, facilitating qualitative and conceptual understanding
based on this foundation.

(6) When the acceleration duration $\delta s$ is relatively short,
specifically $\delta s\ll\frac{1}{a}\sim\frac{1}{T_{U}}$, shorter
than the characteristic time scale corresponding to the formation
of the Unruh thermal equilibrium temperature $T_{U}$, the system
cannot achieve thermal equilibrium within such a brief period. Consequently,
the spectrum of the accelerated point particle does not evolve into
a maximally mixed blackbody spectrum in this short time but only slightly
deviates from a quantum pure state, exhibiting a slight broadening
of its spectrum. This effect can be regarded as a local short-time
extension of the Unruh effect, which typically arises from a long-time
and uniform acceleration in a flat background. This aspect is not
readily apparent from the short-time $\delta s$ perturbation method
employed in this paper, instead, a more rigorous coordinate transformation
is required. For a detailed proof of this point, refer to the author's
previous work \citep{Luo:2023eqf}.

(7) Our starting point is the standard relativistic particle action,
it is the quantum non-inertial effect coming from the second moment
fluctuation that modifies the physical quantities related to the square
or magnitude of acceleration. This marks the difference between our
theory and most other theories \citep{Milgrom:1992hr,Milgrom:2011kx,Milgrom:2014uwk,Milgrom:2022ifm,Milgrom:2023pmv,Namouni:2015pga,2017,Costa:2019pbz}
that attempt to explain MOND from the kinematic perspective. Those
theories seek to modify the non-relativistic particle action, while
we explain MOND through the squared acceleration associated in the
relativistic particle action. Further more, in the sense of the quantum
equivalence principle, there is no essential difference between the
modified-kinematics $a^{i}\rightarrow a_{eff}^{i}=\frac{1}{2}\nabla_{i}Z_{eff}$
and modified-gravity $g_{\mu\nu}\rightarrow Z_{eff}g_{\mu\nu}$ or
$\Phi\rightarrow Z_{eff}\Phi$, contrary to the usual MOND interpretation.
It is for these reasons, the time non-locality constraint \citep{Milgrom:1992hr,Milgrom:2022ifm},
which is considered unavoided in conventional modified-inertia approach
to MOND that are based on modifying the particle kinetic action and
its classical equations of motion, is absent in this framework based
on the quantum non-inertial effect of a standard relativistic particle
action, simply because that our theory does not modify kinematics
at the level of classical equations of motion, but introduces corrections
to the mechanical quantities associated with the acceleration squared
term at the level of second moment quantum fluctuations by the proposed
effect. As a consequence, the difficulties associated with the time
non-locality, such as the incompatibility with local causality of
relativity, are also absent in the present theory.

(8) The enigmatic possible connection between the constant $a_{0}$
fitted in the baryonic Tully-Fisher relation and the square root of
the cosmological constant $\sqrt{\Lambda}$ or the Hubble constant
$H_{0}$ has long remained a mystery. The theory presented in this
paper offers a potential cosmological origin for the constant $a_{0}$,
namely $a_{0}\approx2\beta^{2}a_{bg}=\beta^{2}\sqrt{\frac{\Lambda}{3}}\approx\frac{\sqrt{\Lambda}}{5.2}$.
Additionally, this paper proposes a possible link between the accelerated
expansion of the universe (dark energy) and the anomalous acceleration
observed at the outskirts of isolated galaxies (dark matter). Specifically,
both phenomena are attributed to the effect of the second-order moment
broadening of cosmological spectral lines. However, for test particles
at the outskirts of isolated galaxies (such as 21cm spectral line),
the second-order moment broadening arises from both their proper radial
acceleration and the background acceleration. In contrast, for test
particles associated with the accelerated expansion of the universe
(such as the spectral lines of Type-Ia supernovae), the broadening
is solely due to the background acceleration (as their proper motion
becomes negligible at sufficiently high redshifts).

(9) The origin of acceleration discrepancies lies in the importance
of the intrinsic second moment quantum fluctuations in the deSitter
universe, which makes the mean value of the derivative (the effective
acceleration $a_{eff}$ (\ref{eq:a_eff})) not equal to the derivative
of the mean value (the first moment acceleration $a_{N}$ given by
the unmodified Newtonian gravity). This explains why in the new kinematic
interpretation, the old kinematic concept $a_{N}$ still holds a strict
and fundamental status (in the interpolating function), the reason
is that it acts as the first moment acceleration corresponding to
the unmodified gravity. But strictly speaking, the unmodified first-moment
Newtonian acceleration (or the related Kepler type rotational velocity,
etc.) has no real physical meaning, unless it is dominate when the
background acceleration $a_{bg}$ or the intrinsic second moment of
the deSitter background is ignorable (e.g. in the solar system or
inside galaxies with relatively strong gravity). The actually observed
kinematic velocities/dispersion and accelerations are all inevitably
associated with the effective accelerations (\ref{eq:a_eff}) that
have been corrected by the second moment, according to the quantum
equivalence principle. The second moment broadening effect of spectral
lines under short-time acceleration or within a deSitter background,
as presented in this paper, although extremely small and difficult
to directly measure in conventional laboratory settings, becomes significant
in the universe, a deSitter spacetime dominated by a positive cosmological
constant. At large redshifts, this second-order moment effect notably
modifies the squared term in the series expansion of the distance-redshift
relation, making the relation deviates from the linear Hubble's law,
which is way we find the expansion is accelerating. Alternative way
is through more precise measurements of spectral line broadening profiles
at high redshifts, which is more difficult. Furthermore, given the
fundamental nature of this kinematic effect and the fact that most
cosmological observations involve spectral line measurements, it is
likely that this effect could influence virtually almost all cosmological
measurements related to squared terms involving spectral lines (or
redshifts). Therefore, if this effect is valid, it may have a broad
impact on precision cosmological measurements and anomalies, such
as the Hubble tension (roughly speaking, for the supernovae measurements
in the local universe, we infer the first derivative Hubble rate using
multiple first moments of distance and redshift data points, which
corresponds to the rate corrected by the second moment of redshift;
while for the CMB measurement in the early universe, if the snapshot
angular diameter distance measured at the single redshift of the last
scattering are merely an underestimated and first moment result without
the second moment correction, like the luminosity distance without
the second moment correction in (\ref{eq:rdistance-redshift}), so
the current Hubble rate inferred from them may also turn out to be
an underestimated or first moment result. Therefore, the tension between
them may originate from the second moment contribution to the difference
between the the first derivative and the first moment which may preserve
the agreement with other canonical constraints). We anticipate that
the widespread second moment effect will be either confirmed or refuted
through future precision cosmological observations. 

At present, it remains unclear whether we can directly measure the
pure first moment acceleration (not the effective acceleration (\ref{eq:a_eff}))
that is completely unaffected by the intrinsic deSitter cosmological
second moment effects, even when it is comparable or smaller than
the background acceleration $a_{bg}$. It is also unclear that the
high-precision astrometric missions such as the Gaia satellite has
to potential to do that, despite it is capable of tracing the proper
motion variations of wide binaries, rather than based on the averaged
and static virial equilibrium. If future technological advancements
allow us to directly measure the pure first moment acceleration unaltered
by the intrinsic second-order effects in cosmolgy, and a distinct
discrepancy can be observed between the mean squared value of the
first moment acceleration and the squared mean value of the effective
acceleration inferred from the virial equilibrium, which may serve
as a test for our hypothesis. Up to now, there is still ongoing academic
debate over whether the existing observational data from the Gaia
satellite supports MOND, in which the statistical second moment fluctuation
of the data points plays a crucial role in determining whether the
observational data tend to be consistent with Newtonian dynamics or
favor MOND \citep{Banik:2023pbo}. Of course, if we can directly measure
the spectral line broadening caused by the acceleration, that will
constitute a more direct piece of evidence, which is independent of
the interpretation of MOND, and it possesses its own independent significance.

(10) The acceleration interpolation relation (\ref{eq:acc_composition}),
has been attempted to be derived under various assumptions. For instance,
it has been proposed based on the effect of the assumed absoluteness
of acceleration at the quantum level (acceleration relative to an
absolute quantum vacuum or aether) \citep{Milgrom:1998sy}; or through
the Unruh temperature effect of acceleration or the Deser-Levine temperature
effect in a deSitter background \citep{Deser:1997ri}; or by considering
the effect of the breakdown of coordinate covariance in the deep-MOND
regime \citep{Milgrom:2019rtd}; or by invoking the violation of the
equivalence principle at the quantum level \citep{Smolin:2017kkb},
where the deviation of the modification factor $Z_{eff}$ in front
of the metric from unity is interpreted as the inequality between
inertial and gravitational masses. In other words, these attempts
have sought to derive MOND at the expense of violating two cornerstones
of fundamental physics: the covariance and the equivalence principle.
However, the interpretation presented in this paper does not require
any violation of coordinate covariance or the equivalence principle.
Here, acceleration is not defined relative to any absolute inertial
frame but merely relative to a co-accelerating/free-falling ``inertial
observer'' in the deSitter universe for whom the metric can be considered
flat shown in (\ref{eq:accelerated action}). Our derivation is based
on introducing acceleration and a deSitter background into a covariant
free-particle action (the $Z_{eff}$ factor) via a general coordinate
transformation, which does not explicitly violate covariance. And
in the interpretation framework of this paper, the equivalence principle
is not violated but rather extended to the quantum level: acceleration,
gravity (curvature), and spectral broadening are equivalent, the classical
equivalence principle establishes the equivalence of the first two,
while this paper proposes to extend this equivalence to include the
equivalence between the spectral broadening and the first two, and
the second moment modified effective acceleration is a real acceleration
manifested in the virial equilibrium.

(11) As the foundation of quantum spacetime theory and quantum gravity,
the quantum equivalence principle cannot be derived from some deeper
microscopic mechanisms. Instead, it serves as the fundamental postulate
and starting point of the theory, making geometrical or physical quantities
modified by the universal quantum second moments, such as the effective
distance quadratic form, effective curvature, and effective acceleration,
behave exactly like real geometrical or physical quantities. The paper
does not attempt to find ``microscopic mechanisms'' that used to
modify the macroscopic virial theorem or the macroscopic virial potential
energy, but in the opposite manner: they can be effectively derived
from the quantum equivalence principle. The validity of the quantum
equivalence principle can only be verified or falsified through experimental
measurements. This point is highly analogous to the situation in the
birth of the relativity: researchers attempted to find explanation
of the constancy of the light speed, for instance via some microscopic
mechanical mechanism of Lorentz contraction for material rulers, but
the correct approach is exactly the opposite: taking the constancy
of the light speed as a fundamental postulate, and deriving the Lorentz
contraction for the universal spacetime from it. 
\begin{acknowledgments}
This work was supported in part by the National Science Foundation
of China (NSFC) under Grant No.11205149, and the Scientific Research
Foundation of Jiangsu University for Young Scholars under Grant No.15JDG153.
\end{acknowledgments}

\bibliographystyle{unsrt}

\end{document}